\documentclass[%
 aip,
 amsmath,amssymb,
 reprint,%
]{revtex4-1}

\usepackage{graphicx}
\usepackage{dcolumn}
\usepackage{bm}

\usepackage[utf8]{inputenc}
\usepackage[T1]{fontenc}
\usepackage{mathptmx}
\usepackage{etoolbox}
\usepackage{xcolor}
\usepackage{soul}

\newcommand{\dl}[1]{\textcolor{black}{#1}}

\newcommand{\riccardo}[1]{\textcolor{black}{#1}}

\makeatletter
\def\@email#1#2{%
 \endgroup
 \patchcmd{\titleblock@produce}
  {\frontmatter@RRAPformat}
  {\frontmatter@RRAPformat{\produce@RRAP{*#1\href{mailto:#2}{#2}}}\frontmatter@RRAPformat}
  {}{}
}%
\makeatother
\begin{document}

\preprint{AIP/123-QED}

\title[]{Liquid water under vibrational strong coupling: an extended cavity Born-Oppenheimer molecular dynamics study}
\author{Jessica Bowles}
\affiliation{Laboratoire de Chimie Théorique, Sorbonne Université, Paris, France.}

\author{Jaime De La Fuente Diez}
\affiliation{{Laboratory CPCV, Department of Chemistry, \'Ecole Normale Sup\'erieure, PSL University, Sorbonne Universit\'e, CNRS, Paris, France}}

\author{Damien Laage}
\affiliation{{Laboratory CPCV, Department of Chemistry, \'Ecole Normale Sup\'erieure, PSL University, Sorbonne Universit\'e, CNRS, Paris, France}}

\author{Johannes Richardi}
\affiliation{Laboratoire de Chimie Théorique, Sorbonne Université, Paris, France.}

\author{Rodolphe Vuilleumier}
\affiliation{{Laboratory CPCV, Department of Chemistry, \'Ecole Normale Sup\'erieure, PSL University, Sorbonne Universit\'e, CNRS, Paris, France}}

\author{Riccardo Spezia}
\email{riccardo.spezia@sorbonne-universite.fr.}
\affiliation{Laboratoire de Chimie Théorique, Sorbonne Université, Paris, France.}


\begin{abstract}
A computational study of liquid water when the system is coupled with a (model) Fabry-Perot cavity is reported. At this end,  the Cavity Born-Oppenheimer Molecular Dynamics approach proposed recently (Li et al., Proc. Nat. Acad. Sci. USA, 2020, {\bf 117},
18324--18331) is employed and different properties of water under vibrational strong coupling (VSC) are investigated. 
Different cavity frequencies are considered, corresponding to different modes in the IR spectrum of liquid water: 
high frequency (corresponding to O--H stretching modes), medium frequency (corresponding to water molecule bending) and low frequencies (corresponding to librational modes). Simulations were done both using classical and quantum nuclear dynamics, this last via Ring Polymer Molecular Dynamics. Similar effect\dl{s} of the cavity are obtained in both cases.
Namely, whereas the infrared spectrum is \dl{observed to be split} for all cavity frequencies, no effect\dl{s} 
on structural properties are \dl{detected}. \dl{In addition,} transport and dynamical properties, including the diffusion coefficient, \dl{molecular reorientation} and hydrogen bond (HB) jump exchange times\dl{,} show no effect \dl{due to cavity coupling} when an extended statistical analysis is performed. 
\end{abstract}

\maketitle

%

\section{Introduction}
In recent years, there has been a growing interest in vibrational strong coupling (VSC) and in particular on its impact on chemical reactivity.~\cite{nagarajan2021chemistry,hirai2023molecular, bhuyan2023rise} 
VSC is obtained by placing a molecular system inside an optical cavity such as a Fabri-Pérot cavity that is resonant with a selected vibrational mode of the matter, thereby generating  a hybrid light-matter state.
When the exchange of (virtual) photons is faster than their dissipation, the strong coupling regime is achieved.

Early approaches focused on the coupling of optical cavity modes with electronic states.~\cite{hutchison2012modifying} However, subsequent research demonstrated the feasibility of coupling molecular vibrations with cavity modes in the infrared (IR) range, a phenomenon referred to as vibrational strong coupling (VSC).~\cite{shalabney2015coherent,simpkins2015spanning}

When the frequency of the cavity ($\omega_c$) is at resonance with the vibrational frequency ($\omega_m$) of the molecular system
then a Rabi splitting ($\hbar \Omega_R$)  is obtained in the infrared (IR) spectrum.
It has been shown that VSC 
can also influence chemical reactivity.~\cite{nagarajan2021chemistry,ahn2023modification, hirai2021selective} 
Specifically, it has been observed that kinetics are altered, as highlighted in different studies,~\cite{thomas2016ground,vergauwe2019modification,lather2019cavity,hirai2020modulation,lather2020improving,hiura2019vacuum,lather2022cavity,Ahn2023,galego2019cavity,li2021theory,climent2020sn} and more recently, some equilibrium properties have been found to undergo modifications. 
These include changes in  conductivity,~\cite{fukushima2022inherent,fukushima2023} and the equilibrium constant in the formation of charge-transfer complexes.~\cite{Pang2020} 
On the other hand, dielectric constant is not affected by VSC, as recently deeply discussed by Michon and Simpkins.~\cite{Michon2024}

While the Rabi splitting in IR spectrum is now well understood,~\cite{EbbesenACR2016,Ahn2018,Hernandez2019,Scholes2021,GeorgePRL2016} it is much less clear how these other properties are modified in VSC regime. 
For this reason, the last years have been characterized by an important theoretical activity in the field.~\cite{mandal2023theoretical, li2020cavity,lindoy2022resonant,yang2021quantum,schafer2018ab,sidler2023numerically,sun2022suppression,yu2022multidimensional}
The general theoretical description of matter under VSC is typically based on the Pauli–Fierz (PF)  quantum electrodynamics (QED) Hamiltonian:~\cite{flick2017atoms}

\begin{equation}
    \hat{H} = \hat{H}_M + \hat{H}_F
\end{equation}
where $\hat{H}_M$ is the  Hamiltonian of the molecular system and 
$\hat{H}_F$ is the interaction with the vacuum field, which can be written as:~\cite{sidler2020polaritonic,sidler2023unraveling, sun2022suppression,fiechter2023quantum,ying2023resonance,mandal2023theoretical,lieberherr2023vibrational} 
\begin{equation}
 \hat{H}_F =  \sum_{k,\lambda}\frac{1}{2}\hat{p}_{k,\lambda}^2 +\frac{1}{2}\omega_{k,\lambda}^2 \left( \hat{q}_{k,\lambda} + \sum_{s=1}^N \frac{ \hat{ {\vec{\mu} }}_s\cdot \vec{\xi}\riccardo{_\lambda}}{\omega_{k,\lambda}\sqrt{V \epsilon_0}}\right)^2,
    \label{eq:PF_ham}
\end{equation}
%
where $k$ denotes the (virtual) photon with given frequency, $\omega_k$, position, $\hat{q}_{k,\lambda}$, and momentum, $\hat{p}_{k,\lambda}$, operators, $\lambda$ the two polarization directions orthogonal to the cavity walls (so typically $x$ and $y$), $V$ the cavity volume, $\epsilon_0$ the vacuum permittivity, $\hat{\vec{\mu}}_s$ the dipole operator for the molecular system, $\vec{\xi}$ the polarization direction of the (virtual) photons and $N$ the number of modes interacting with the cavity field.

Various theoretical studies using this Hamiltonian have been performed in recent years.
A first kind of approach focused on adapting  transition-state theory (TST) for the presence of the 
cavity and thus obtained rate constants with an effective friction, in analogy with Kramers and Grote-Hynes theories.~\cite{kramers1940brownian,grote1980stable}
They generally considered only one mode for the matter coupled to the cavity with effective coupling parameters.
For example, Li et al.~\cite{li2021theory} applied this theory to explain changes in the reaction rate under strong coupling.
Also Lindoy et al.\cite{lindoy2022resonant} used an analytical rate  which shows that chemical kinetics can be modified (enhanced or suppressed) due to a modified friction under VSC. 
A quantum description via TST of cavity-catalysed adiabatic chemical reaction was then carried out by  Yang and Cao,
\cite{yang2021quantum} providing an explanation for the maximum modification when the cavity frequency is near the vibrational resonance frequency and also for the collective and selective aspects.

In a similar vein, quantum dynamics were performed using different
models, but generally considering only one matter mode coupled with the
cavity. Morse dissociation or double well models were studied in this
way, showing an effect on the rate.~\cite{lindoy2023quantum,Mondal2022}
Also quasi-classical trajectories were employed to reveal the 
modification of water dimer dissociation due to VSC.~\cite{YuBowman2023}
Notably, using the open quantum system
formalism, Reichmann and co-workers showed that it is possible to
obtain the highest enhancement when the cavity is in resonance with
the reactant mode.~\cite{lindoy2023quantum}
%

Another approach, developed in Rubio's group, is  based on the application of the coupling to the full nuclear-electron Hamiltonian.~\cite{schafer2018ab}
This method was used, for example, to examine whether the collective coupling  implies local modifications of chemical properties~\cite{sidler2020polaritonic} and applied on a HD$^+$ molecule.~\cite{sidler2023numerically}
%
Furthermore, quantum dynamics have been employed to describe a model of a chemical reaction in the condensed phase within the cavity.~\cite{lindoy2023quantum, fiechter2023quantum,ying2023resonance}
However, these methods based on quantum dynamics 
only allow the description of model systems and most of the time, the matter is described with only one explicit mode.

To study how VSC modifies the properties of a condensed phase system, an approach based on molecular dynamics simulations was 
developed by Subotnik and co-workers,~\cite{li2020cavity} called Cavity 
Molecular Dynamics (CavMD). In this approach, the atoms of the system interact with classical force fields and the
cavity acts on the dipole of the molecules using a simple point charge representation of the molecular charge distribution (as in classical molecular dynamics).
This approach was first used to study liquid water \cite{li2020cavity}
and then extended to other systems. 
For example, it is possible to study the 
nonlinear response of carbon dioxide under VSC,~\cite{li2021cavity} revealing that a strong pulse can lead to an enhanced overall molecular nonlinear absorption that causes an ultrashort  lifetime (0.2 ps) of the lower hybrid light–matter state.
Furthermore, 
it has been employed to show that intermolecular vibrational energy transfer is facilitated under VSC.~\cite{li2021collective}
The same authors also numerically illustrate that IR pumping a liquid confined in an optical cavity under vibrational strong coupling can  selectively excite a small concentration of solute molecules.~\cite{li2022energy} 
 Notably, this approach can be extended to include phenomenologically the cavity loss.~\cite{li2022polariton}
The CavMD approach has also been applied, using a QM/MM approach, to study Fe(CO)$_5$ in n-dodecane \cite{li2022qm} predicting a preferred vibrational energy transfer to the equatorial CO vibrations rather than the axial CO vibrations when exciting the lower polariton.

Calculations on liquid water done by Subotnik and co-workers where done setting the cavity at resonance with the O--H vibrational mode and
they do not observe any relevant effect on properties like water structure, diffusion coefficient or dielectric constant. 
However
water has different vibrational modes and our main interest is to see if a cavity in resonance with these modes can be at the origin
of the modification of some properties.
At this end we have considered a simple molecular dynamics simulation of liquid water 
using the approach proposed by Li et al.,~\cite{li2020cavity} and we investigated different properties as the cavity is tuned at different  frequencies. 
%
Notably, based on the works of Laage, Hynes and co-workers, 
it is known that many dynamical processes in liquid water, including, \textit{e.g.}, translational diffusion and reorientation, are governed by hydrogen-bond exchanges.\cite{laage2006molecular,laage2008molecular,laage2011reorientation,wilkins2017nuclear,Gomez2022} The rate-limiting step of these hydrogen-bond exchanges has been shown to be the combination of the initial hydrogen-bond elongation and of the new hydrogen-bond contraction. 
This motion is  a low-frequency motion and in previous theoretical studies on 
VSC on liquid water, the cavity frequency was set only on high frequency values.~\cite{li2020cavity,li2022quantum} 
Here, we have tuned the cavity frequency in different spectral regions corresponding not only to O--H stretching but also to the bending and to the vibrational modes. 
Furthermore, Sun and Rubeiro\cite{sun2024theoretical} have pointed out that coupling the cavity with modes at lower frequency ($\sim$500-600~cm$^{-1}$) has a higher impact on an S$_N$2 reaction.
We have thus analyzed if different properties are modified under VSC. In particular, dynamical properties, related to hydrogen bond dynamics were investigated as well as transport properties.
We are motivated  by experimental studies showing that diffusion coefficient of water can be modified under VSC.~\cite{fukushima2022inherent,fukushima2023} Since also low vibrational
frequencies can be related to water diffusion (e.g. through the aforementioned H-bond exchange mechanism), we have extended the
study of liquid water under VSC using the CavMD approach to other  bands present in the water IR spectrum.

In the first part, we restricted ourselves to classical evolution of particles, in order to strengthen the importance of a large statistical sampling to minimize the uncertainties on the calculated properties. 
It was shown that using Path-Integral based MD, the IR splitting changes slightly, compared to classical simulations, as well as other properties like  
the H-bond jump time.~\cite{lieberherr2023vibrational,li2022quantum,wilkins2017nuclear} 
Thus in a final part ring-polymer molecular dynamics (RPMD) was employed to study nuclear quantum effects on the properties.~\cite{craig2004quantum}


The remainder of the article is as follows: first we will describe the basic principles of the methods and the simulations performed. Then we will
show that polaritonic states are obtained for all the cavity frequencies, and then we will report results on structural and thermodynamic properties then transport and dynamical properties.
A discussion on the statistical sampling needed  in such simulations is presented, followed by the study of possible nuclear quantum effects using RPMD.
The last section provides conclusions and outlook for future studies.

\section{Methods}
\subsection{Molecular Dynamics Simulations}
We simulated a system composed by a box of 216 water molecules using the
flexible q-TIP4P/F force field.~\cite{habershon2009competing} 
The box size was set at 18.64~\AA~to reproduce
the experimental water density at 300~K, corresponding to what used by Li et al.~\cite{li2020cavity} 
Standard periodic boundary conditions with an Ewald summation for the  intermolecular Coulombic interactions were used.
%
In the  classical nuclear dynamics simulations, for each cavity frequency (and for the simulation without the cavity), we first equilibrated the system in the canonical ensemble at 300~K, using a Langevin thermostat with a friction caracteristic time $\tau$~=~100~fs, and a time step of 1~fs, using the velocity Verlet scheme.~\cite{verlet1967computer} 
We then selected 30 initial conditions to run NVE simulations of 100~ps 
with a propagation time step of 0.5~fs to obtain all the properties except the dielectric constant where 20 NVE trajectories of 1~ns were performed and the IR and VDOS where 4 trajectories of 20~ps were carried out. 
In the RPMD simulations, the equilibration trajectories  employed a white noise Langevin  thermostat on the beads and a global velocity rescaling thermostat on the centroid mode, with $\tau$~=~100~fs.~\cite{ceriotti2010efficient,rossi2014remove}
%
\riccardo{For the dielectric constant, we then extracted 50 initial conditions and ran T-RPMD trajectories of 300~ps each, in which only the internal modes} are thermostatted,~\cite{ceriotti2010efficient} using the positions and velocities extracted from the equilibrated trajectories as initial conditions.
For the dynamical properties, 30 NVE trajectories  of 100~ps, also started from the equilibrated trajectories, were exploited when we used 8 beads and just two trajectories of 100~ps for 32 beads. All RPMD simulations employed a time step of 0.25~fs. 

The 
description of the cavity 
 followed the methodology suggested by Li et al \cite{li2020cavity} using the CavMD software,~\cite{CavMD-github} 
 a modified version of  i-PI.~\cite{kapil2019pi}  
This software is coupled here with the LAMMPS package to use classical force fields.~\cite{LAMMPS}
CavMD models the system through two equations of motions, one corresponding to the atoms and one corresponding to the virtual photons of the cavity.
The equations of motion used are obtained after reducing all of the operators in Eq. \ref{eq:PF_ham} to the corresponding classical observables: 
\begin{equation}
   M_j \mathbf{\ddot{R}}_j = \mathbf{F}^{(0)}_j - \sum_{k,\lambda} \left(  \epsilon_{k,\lambda}\tilde{q}_{k,\lambda} +\frac{\epsilon_{k,\lambda}^2}{m_{k,\lambda}\omega_{k,\lambda}^2} \sum_{s=1}^N \vec{\mu}_s\cdot\vec{\xi}_{\lambda} \right)\frac{\partial \vec{\mu}_j\cdot\vec{\xi}\riccardo{_\lambda}}{\partial \mathbf{R}_j} , 
   \label{eq:Eq_motion1} 
\end{equation}
\begin{equation}  
   m_{k,\lambda}\ddot{\tilde{q}}_{k,\lambda} = -m_{k,\lambda}\omega_{k,\lambda}^2\tilde{q}_{k,\lambda} - \epsilon_{k,\lambda} \sum_{s=1}^N \vec{\mu}_s\cdot\vec{\xi}\riccardo{_\lambda}, 
   \label{eq:Eq_motion2} 
\end{equation}
\begin{equation}
 \text{ with the coupling  constant: } \;\;  \epsilon_{k,\lambda}=\sqrt{m_{k,\lambda}\omega_{k,\lambda}^2/V \epsilon_0}.
 \label{eq:coupling_cste}
\end{equation}
In Eq. \ref{eq:Eq_motion1} and \ref{eq:Eq_motion2}, an auxiliary mass $m_{k,\lambda}$ is introduced to align with standard MD simulations where particle mass information is required. 
Consequently, the positions representing the cavity are modified as follows: $q_{k,\lambda} =\tilde{q}_{k,\lambda} \cdot \sqrt{m_{k,\lambda}}$.

The positions and velocities of the atoms obtained from the dynamics are used to determine the properties of interest. 
Various values of the cavity frequency are chosen that correspond to frequencies where a signal is present in the IR spectrum, such as 3550, 1650, 600 and 400~cm$^{-1}$. For each cavity frequency, a different value of the coupling constant is chosen, 
to have the same interaction strength with the cavity for each frequency value, $\epsilon_{k,\lambda}\omega_{k,\lambda}=\mathrm{cte}$. 
For the cavity frequency at 3550~cm$^{-1}$  the same coupling constant is chosen as in Li et al.~\cite{li2020cavity}: $4\cdot10^{-4}$ a.u. then the other values are adapted linearly from this following Equation~\ref{eq:coupling_cste}: $1.9\cdot10^{-4}$, $6.8\cdot10^{-5}$, $4.5\cdot10^{-5}$ a.u. respectively.

\riccardo{As pointed out by Li et al.~\cite{li2020cavity}, for lower cavity modes overtones of the cavity should also be included. We have then run
additional trajectories setting the cavity fundamental frequency at 600~cm$^{-1}$, and including additional cavity overtone modes at 1200, 1800, 2400, 3000 and 3600~cm$^{-1}$ using the multi-cavity method described by Li et al.~\cite{li2020cavity} and for two values of coupling strength of $2.0\cdot10^{-4}$ and $6.8\cdot10^{-5}$ a.u.} .

We should point out that while the water model used does not reproduce accurately all the desired liquid water properties, it has the great advantage of being widely used and is one of the simplest flexible water models. It was used by Li et al.~\cite{li2020cavity} in their first study of water under VSC and thus it is a natural choice for an initial exploration of the impact of VSC on dynamical properties and with different cavity frequencies. We emphasize that our focus will be not much on the absolute values of the different properties, but rather on how (and if) they 
evolve under VSC.

\subsection{Calculations of Properties}
From MD simulations we directly obtained different properties such as the infrared spectrum, the vibrational density of states (VDOS), the radial distribution function (RDF), the dielectric constant, the diffusion coefficient, the H-bond reorientation time and the H-bond jump times. 
Here we list briefly how they are obtained, while in the original articles more details are reported.

\begin{enumerate}
    
\item The infrared spectrum is determined as the Fourier transform of the auto-correlation function of the total electric dipole moment of the molecular system and  the VDOS from the auto-correlation function of the velocities. As in the model, and shown in the original paper by Li et al.,~\cite{li2020cavity} the direction of the cavity lies along the $z$-axis, 
so the IR spectra along $x$- and $y$-axis are those who should show the formation of a polaritonic state. 

\item The dielectric constant, $\varepsilon$, is calculated from the usual equation for tin foil periodic boundary conditions with Ewald sums (dielectric constant of surrounding is $\infty$):~\cite{allen2017computer,richardi2005fast,Leeuw1980}
\begin{equation}
    \varepsilon = 1 + 3 y_D g_K
\end{equation}
where the Kirkwood factor $g_K= (\left<M^2\right>-\left<M\right>^2)/ \left( N\mu^2 \right)$ and the dipole parameter $y_D=\rho \mu^2 / 9k_B T \varepsilon_0 $. 
$N$ is the number of molecules, $\mu$ is the absolute value of the molecular
dipole averaged for each molecule and $M=\sum_i\vec{\mu}_i$,  $\rho$ is the number density and $\varepsilon_0$ is the vacuum permittivity.
Due to the well known slow  convergence of the dielectric constant, we calculated that on longer trajectories (20~ns). 

%
\item The diffusion coefficient~\cite{zhang2017second} is obtained through the slope of the mean square displacement (MSD) of the oxygen atoms. 
Here we limited ourselves to the diffusion coefficient corresponding to a given water box length, since we are interested into the impact of VSC keeping the same simulation conditions. In fact, it is well known that hydrodynamics suffers finite size effects and to recover the diffusion coefficient one should extrapolate the infinite box value.~\cite{yeh2004system}
%

%

\item To determine the reorientation time, the angle between the orientation of an OH bond at time $t_0$ and $t_1$ is examined.
The  order 2  Legendre polynomial of the cross correlation between these two vectors $P_2 [v(t_0) \cdot v(t_1)]$ is determined. 
Then the natural logarithm of the Legendre polynomial is fitted 
over a time interval of 10~ps to obtain the slope as $ln(P_2)=- t/\tau_{reor}$.

\item To determine the jump time, 
the cross-correlation function of the probability for a water OH group to donate a stable hydrogen-bond to an initial acceptor $I$ at time 0, $n_I(0)$, and to form a hydrogen-bond to a new stable partner $F$ at time $t$, $n_F(t)$, is used:~\cite{laage2008molecular}
\begin{equation}
    C_{I,F}(t) = \langle n_{I}(0) n_{F}(t) \rangle.
\end{equation}
\dl{Here, $n_{I,F}(t)=1$ when the OH group under consideration donates a stable H-bond to $I,F$ at time $t$ and $n_{I,F}(t)=0$ otherwise. The H-bond presence is determined from strict geometric criteria ($R_{OO}<2.95$~\AA, $R_{HO}<2.05 t$~\AA, and $\theta_{HOO}<20^\circ$)\cite{laage2008molecular}.} 
This correlation function gives the probability to form a stable final \dl{H-bonded} state at time $t$ when the system was \dl{engaged in a stable H-bond with a different acceptor} in the initial state at time $0$. \dl{Absorbing boundary conditions in the product state are applied to consider the first acceptor exchange only.} Thus the decay
time of the complementary probability, $1 - C_{I,F}(t) ) = exp(-t/\tau_{jump})$, provides the jump time $\tau_{jump}$ corresponding to the time for a water molecule to switch \dl{stable H-bond acceptors}.~\cite{laage2008molecular}

\item Finally, it was shown that the free energy barrier corresponding to an hydrogen bond exchange \dl{can be estimated} from
the oxygen-oxygen radial distribution function.~\cite{wilkins2017nuclear} This \dl{free energy} barrier \dl{is the sum of two terms, which correspond respectively to the} elongation of the distance between the oxygen atom of the jumping H-bond donor water and the initial acceptor molecule, and \dl{to the} compression of the distance between the oxygen atom of the same H-bond donor molecule \dl{and} the new acceptor molecule. \dl{Each term can be determined from the potential of mean force along the oxygen-oxygen distance}, $W(r)$, which is 
related to the oxygen-oxygen radial distribution function (RDF), $g(r)$: 
\begin{equation}
    W(r) = - k_B T \ln [g(r)]
\end{equation}
where $g(r)$ corresponds to the O-O RDF. We denote $r_{max1}$, $r_{max2}$ and $r_{min}$ the distances corresponding to the first two maxima and the first minimum respectively of the RDF. 

In the reactant state the average oxygen-oxygen distance corresponds to the first maximum of the $g(r)$ ($g(r_{max1})$, a minimum in the $W(r)$) and, 
before the jump, the oxygen atom of the final state will be at an average separation corresponding to the second maximum in the $g(r)$ ($g(r_{max2})$, a 
second minimum in the $W(r)$). 
It is then possible to estimate the free energy barrier of the H-bond exchange transition state from $g(r)$ calculated from simulations as:
\begin{equation}
  \Delta G^{\ddagger} = -k_B T \left[ \ln\left(\frac{g(r_{min})}{g(r_{max1})}\right) + \ln\left(\frac{g(r_{min})}{g(r_{max2})}\right)\right].
  \label{eqn:Energy_barrier}
\end{equation}

%
Thus, we estimated the free energy barrier for the hydrogen-bond switch reaction simply from the RDF calculated from the simulations as
a function of the cavity frequency. 
\end{enumerate}

In all the article, the error bars are reported as a confidence interval of 95\%. 

\section{Results and Discussion}
\subsection{Polariton Formation}

\begin{figure}
    \centering
    \includegraphics[width=0.7\linewidth]{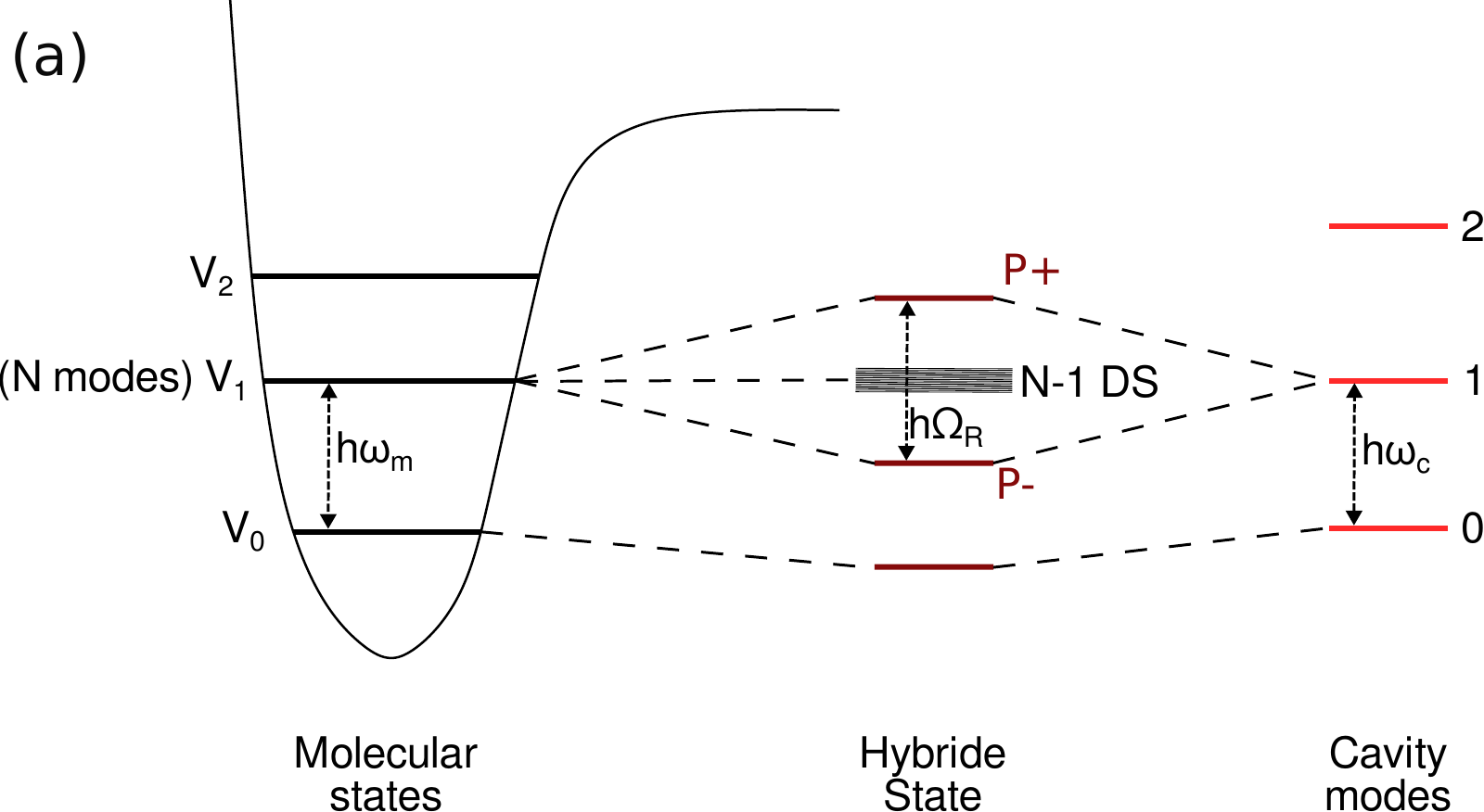} \\
    \vspace{0.4cm}
    \includegraphics[width=0.9\linewidth]{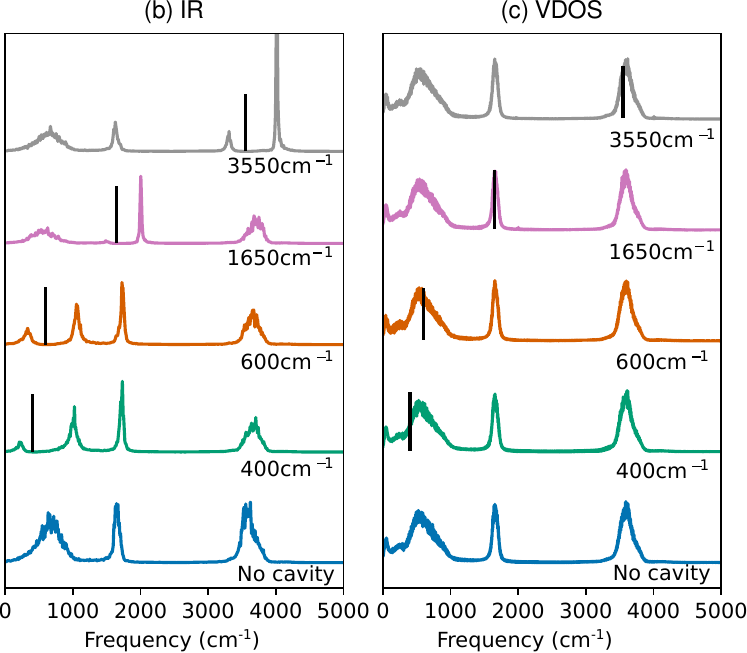}
    \caption{(a) Scheme of vibrational strong coupling (b) IR spectra with cavity frequencies of 400, 600, 1650, 3550~cm$^{-1}$ and excluding the cavity (c) VDOS spectra with the same cavity frequencies. The black vertical lines represent the cavity frequency and the positions and velocities were collected over 80~ps. } 
    \label{fig:IR-VDOS_single}
\end{figure}

As previously discussed, the CavMD approach is able to mimic the formation of a 
hybride state in which the vibrational modes of liquid water are
mixed with the resonant cavity mode (see a pictorial explanation in Figure~\ref{fig:IR-VDOS_single}a).
The IR spectrum of liquid water, obtained from 4 NVE trajectories with the q-TIP4P/F model was calculated from the Fourier transform of the dipole-dipole correlation function on the (x,y) plane.
In the absence of cavity the IR spectrum exhibits
 the well-known  bands (as shown in the bottom of Figure~\ref{fig:IR-VDOS_single}b) corresponding to the librations
(around 600~cm$^{-1}$), the bending mode (at 1650~cm$^{-1}$) and OH stretching (at 3550~cm$^{-1}$). 
%
We chose to turn the cavity to each of these modes succesively.
Since the librational band is quite broad, we have also set the cavity to a lower value, 400~cm~$^{-1}$.
Results are summarized in
Figure~\ref{fig:IR-VDOS_single}b.


When the cavity frequency is at resonance with a water vibrational mode, a splitting can be seen in 
the resulting IR spectrum, corresponding to the formation of two polaritonic modes.
The two new peaks correspond to the P$^+$ and P$^-$  polaritonic modes that arise from the hybrid state as shown \riccardo{in the IR spectra reported in} Figure~\ref{fig:IR-VDOS_single}(b). 
On the other hand, for the VDOS spectra, shown in Figure~\ref{fig:IR-VDOS_single}(c), no splitting occurs, 
since no selection rules apply and so also the modes which are not hybridized with matter, the dark states (DS), are present.
These states are numerically dominant over the
bright states as depicted in
Figure~\ref{fig:IR-VDOS_single}(a) and explained numerically in details by de la Fuente et al.~\cite{JaimeAccepted}

\begin{figure}
    \centering
    \includegraphics[width=1\linewidth]{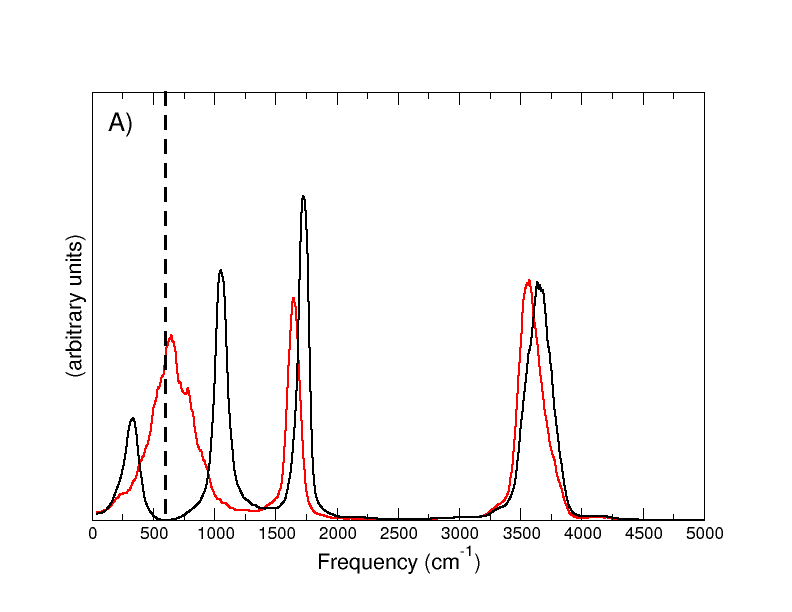} \\
    \includegraphics[width=1\linewidth]{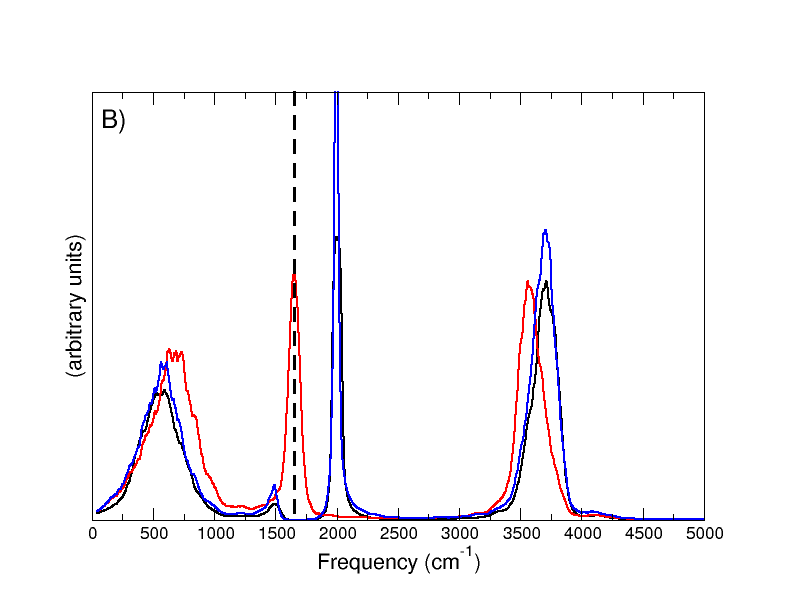} \\
    \includegraphics[width=1\linewidth]{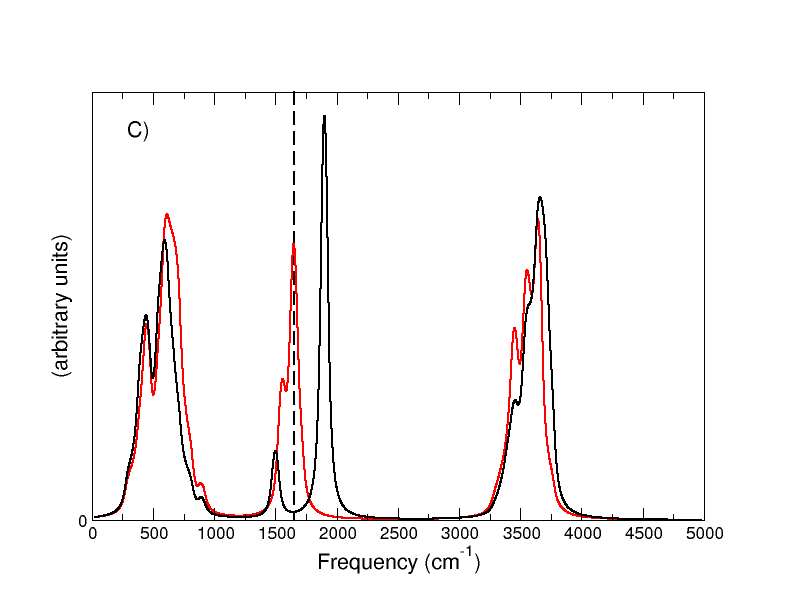} \\
    \caption{IR spectra under VSC as obtained from signal in the (x,y) plane (black solid line) and z-axis (red line) for two cavity frequencies: 600~cm$^{-1}$ (panel A) and 1650~cm$^{-1}$ (panel B). 
\riccardo{In Panel B we added also the spectrum obtained from the harmonic oscillator model developed by Lieberherr et al.~\cite{lieberherr2023vibrational} (blue line).}
    In Panel C we report the results obtained from a simple 3-mode model with the cavity set at 1650~cm$^{-1}$: \riccardo{red} line the IR without coupling with the cavity and in \riccardo{black} the coupled spectrum. In all panels we report as vertical dashed lines the cavity frequency.} 
    \label{fig:IR-600}
\end{figure}

In Figure~\ref{fig:IR-600} (panels A and B), we show in details the IR spectra obtained under resonance conditions: in particular we show both
the spectrum calculating from dipoles in the (x,y) plane, corresponding
to the plane on which the field is coupled to the matter, and in the z-axis. In panel A, we show the spectra obtained when the cavity is
set to 600~cm$^{-1}$, while in panel B when it is set to  
1650~cm$^{-1}$. 
Interestingly, we notice that for the signals in the light polarization plane, where the polaritons are formed, the other bands, which are out-of-resonance with the cavity, are also shifted. When 
the cavity is in resonance with the librational motion, we observe that the bending is blue shifted by about 80~cm$^{-1}$ (the maximum being
now at about 1730~cm$^{-1}$) and the stretching is also blue shifted by about 100~cm$^{-1}$ (the maximum is now at about 3650~cm$^{-1}$).
The same phenomenon is observed when the cavity is resonant with the bending: now the broad librational broad band is red-shifted, while the 
stretching is still blue shifted. 
This phenomenon is probably due to the fact that the cavity is off-resonance with the other modes, while it remains close enough to
slightly shift them.  
\riccardo{These spectral features, namely the asymmetry of the polaritons and the shifts of the other bands, are obtained also by employing
a simple harmonic oscillator model proposed by Lieberherr et al.~\cite{lieberherr2023vibrational} in which the cavity-free IR spectrum was given as unperturbed input. Results are shown in the same Figure~\ref{fig:IR-600}, panels A and B.}

\riccardo{To understand these particular features of polaritonic bands, we have also} employed a model Pauli-Fierz Hamiltonian where the matter is represented by 
three modes corresponding to the frequencies of liquid water.
To represent the broadening of liquid phase, 
 a distribution of frequencies is considered for each mode, \riccardo{reflecting the distribution width obtained from cavity-free simulations}. The cavity is then
added at a frequency of 1650~cm~$^{-1}$. With this Hamiltonian,  Path Integral Monte Carlo simulations allow  to
obtain IR spectrum and polaritons in very good agreement
with exact quantum Hamiltonian, as shown recently by some of
us.~\cite{JaimeAccepted} More details are reported in Appendix~\ref{Appendix:PIMC}. 

In panel C of Figure~\ref{fig:IR-600}, we report results obtained with this 3-mode model with the cavity set at 1650~cm~$^{-1}$
\riccardo{(note that the non exactly unimodal IR bands of the uncoupled system is due to the numerical sampling of 100 vibrational modes)}. 
Notably, we observe that the high
frequency peak is blue shifted while the low frequency is red shifted. This confirms that the Born-Oppenheimer cavity-MD 
approach is able to correctly reproduce spectroscopic properties
for systems under VSC, even concerning detailed features such as
the shift of modes not directly at resonance with the cavity.

\riccardo{We also obtain that the lower polariton (LP) is much less intense than the upper polariton (UP), as in molecular simulations. 
From PIMC simulations,
we can obtain the coefficients of the cavity and of the matter modes in the coupled system which reflect the light-matter hybridization and
can be used to get deeper details. In particular, we have analyzed how the cavity coefficients are distributed as a function of frequency. In the simple
case of a "isolated" polariton, the cavity contribution is 50~\% on the LP and 50\% on the UP.}~\cite{JaimeAccepted} 
\riccardo{
Here, however, we observe that the cavity 
has a non negligible contribution also away from the polaritonic band. 
Notably we divided the spectrum in four regions: (I) $\omega$~<~1300~cm$^{-1}$; (II) 1300~<~$\omega$~<~1650~cm$^{-1}$; (III) 1650~<~$\omega$~<~2500~cm$^{-1}$
and (IV) $\omega$~>~2500~cm$^{-1}$. 
Regions II and III correspond to LP and UP. From the cavity coefficients obtained from PIMC simulations we got that while the UP (region III) has almost 50\% of the cavity, the LP (region II) has about 35\% of the cavity and region I has about 15\%. This explains the decrease in intensity of the LP.}

\riccardo{Furthermore, from PIMC simulations it is possible to evaluate the degree of delocalization of the different bands by calculating
the Inverse Participation Ratio (IPR) :
\begin{equation}
    IPR_i = \sum_{j=1}^{N} |c_j^{(i)}|^4
\end{equation}
where $c_j^{(i)}$ are the coefficients for each matter mode $j$ of the final band $i$ and $N$ are the number of matter modes. In our
previous study where only one vibrational band was considered, we obtained that the polaritonic bands have very low values, corresponding
 to high delocalization (note that by construction the maximum delocalization is $1/N$, so 0.01 in the present model). Here we obtain that the upper polariton has a IPR$_{UP}$~=~0.012, while the lower polariton has IPR$_{LP}$~=~0.16.}
\riccardo{This means that the upper polariton is a "normal" polariton (i.e. as obtained when studying a single isolated vibrational mode), such that its delocalization is close to the maximal one.}
\riccardo{On the other hand, the lower polariton is more localized, with a value similar to what we obtained for a dark state.~\cite{JaimeAccepted}
It is still bright (but much less intense than UP) because the cavity coefficient is not zero even if lower than for UP because it is spread
over the LP and lower frequencies (the librational band in the present case).}


\subsection{Structural and Thermodynamic Properties}

In the following section, various structural and thermodynamic properties such as the  oxygen-oxygen radial distribution function (RDF), the potential of mean force (PMF), the free energy barrier corresponding to H-bond exchange $\Delta G^{\ddagger}$ and the dielectric constant are examined. 

\begin{figure}
    \centering
    \includegraphics[width=0.9\linewidth]{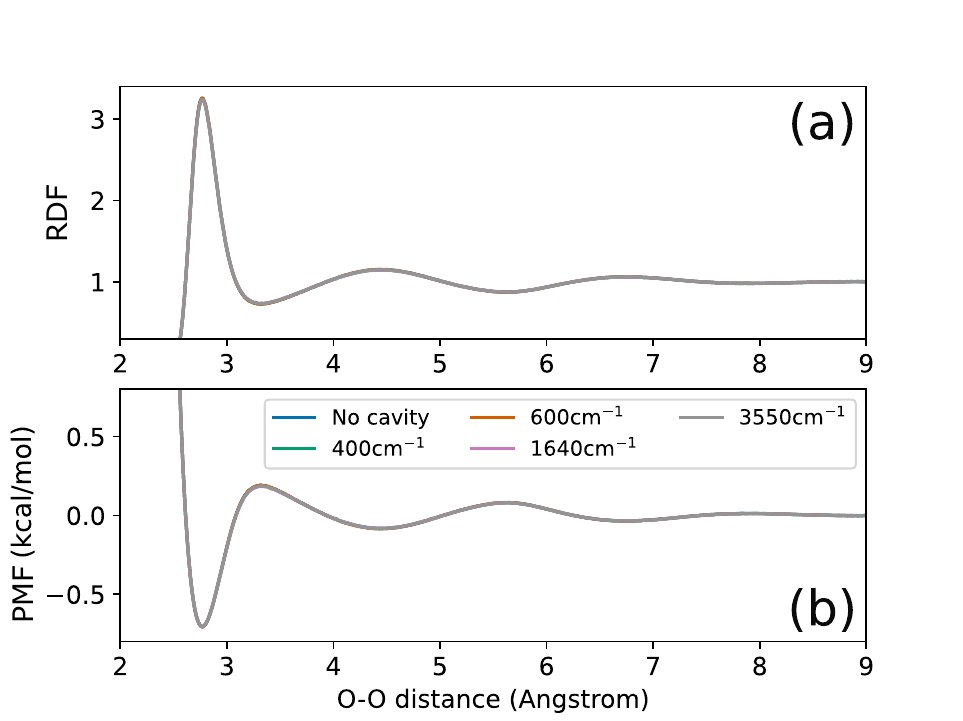}
    \caption{(a) O-O Radial distribution function (RDF) and (b) potential of mean force (PMF). 
    The values without the cavity and for various cavity frequencies are plotted 
    but are indistinguishable since they are all superimposed. 
}
    \label{fig:gr_PMF}
\end{figure}

The water structure is typically probed by the oxygen-oxygen radial distribution function  which is reported in Figure~\ref{fig:gr_PMF}(a). Here we report also the RDFs obtained when the cavity is resonant with different modes  
obtaining almost no effect,
similarly to what was reported by Li et al.~\cite{li2020cavity} for the cavity  set only at 3550~cm$^{-1}$. Note that in the figure
the curves overlap almost perfectly, such that they cannot be distinguished. 

From the RDF, it is possible to estimate the free energy barrier for H-bond exchange, as summarized previously and discussed in details 
by Wilkins et al.~\cite{wilkins2017nuclear}
The barriers as a function of cavity frequency are obtained from the potential of mean force reported in Figure~\ref{fig:gr_PMF}(b).
Also the PMFs are not affected by the presence of a resonant cavity. 
Unsurprisingly, since there is no large change when including the cavity on both RDF and PMF, 
there is also no  effect on the $\Delta G^{\ddagger}$. 
Values are summarized in Table~\ref{tbl:D_tau}. This means that the thermodynamics of H-bond exchange is not modified 
when including a cavity mode with the CavMD formalism.

We should also notice that our values are in agreement with the free
energy barrier obtained previously by Wilkins et al. (1.22~kcal/mol)~\cite{wilkins2017nuclear} with the same water model but with no cavity.


To continue the study of water within the cavity, we look at the dielectric constant, $\varepsilon$, obtained as function of cavity frequency. In this case, \riccardo{the total simulation time is increased from 3~ns to 20~ns} due the well-known slow convergence of this property. 
Results
are summarized in Table~\ref{tbl:D_tau} which also includes values reported in the literature using the same water model.
Also in this case, we cannot see any significant effect on the cavity. The results found in the literature with the same water model (without the cavity or with the cavity set at 3550~cm$^{-1}$) also correspond to our findings. 

The fact that no effect is observed for structural and thermodynamic properties, while searched extensively, should be expected for classical particles under the Pauli-Fierz Hamiltonian, as a consequence of classical statistical mechanics. \riccardo{A concise derivation is given by Li et al.~\cite{li2020cavity} and in the Appendix~\ref{Appendix:Inva-Equi-Prop} we provide a \dl{more extended} derivation.}

\subsection{Transport and Dynamical Properties}

We now dive into transport and dynamical properties of liquid water that can be extracted from molecular dynamics simulations, 
and examine the diffusion coefficient, the water reorientation time and the H-bond jump time. 
As discussed previously, 
to obtain transport and dynamical properties we averaged, for each value of cavity frequency ($\omega_c$) over 30 NVE trajectories generated from a long NVT trajectory. 
The cavity is tuned to the same frequencies as for the structural and thermodynamic properties. 

\begin{figure}
    \centering
    \includegraphics[width=0.9\linewidth]{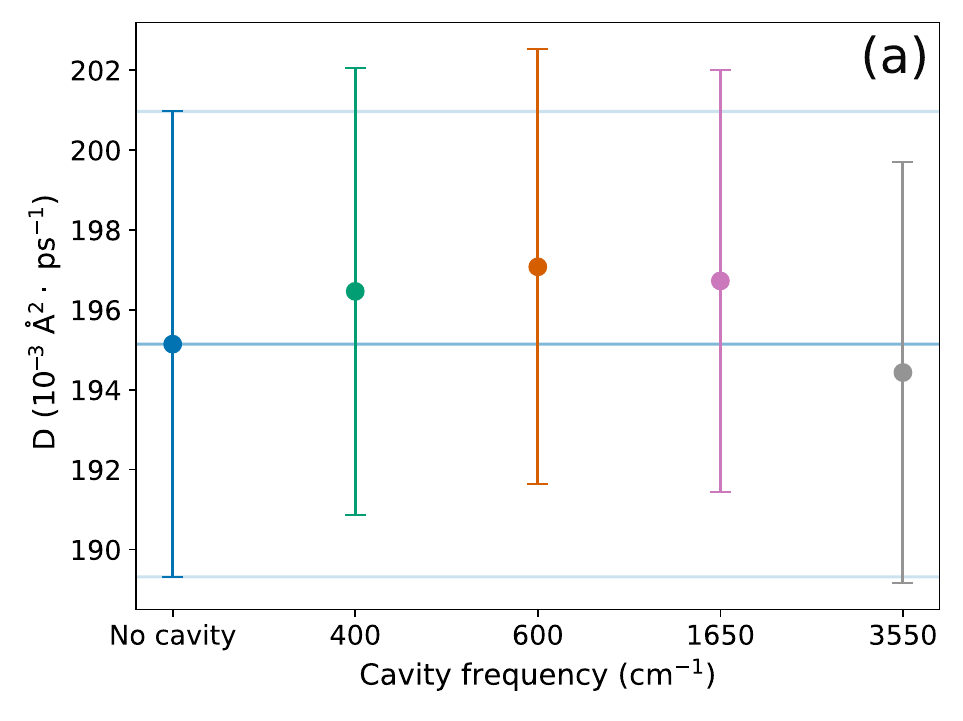} \\
    \includegraphics[width=0.9\linewidth]{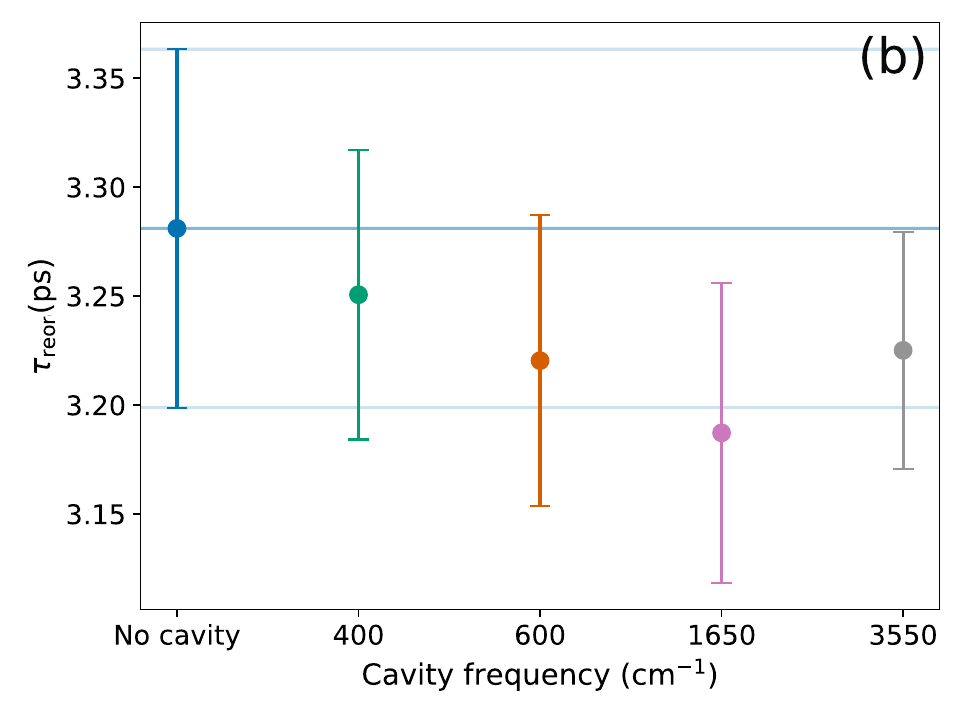}
    \includegraphics[width=0.9 \linewidth]{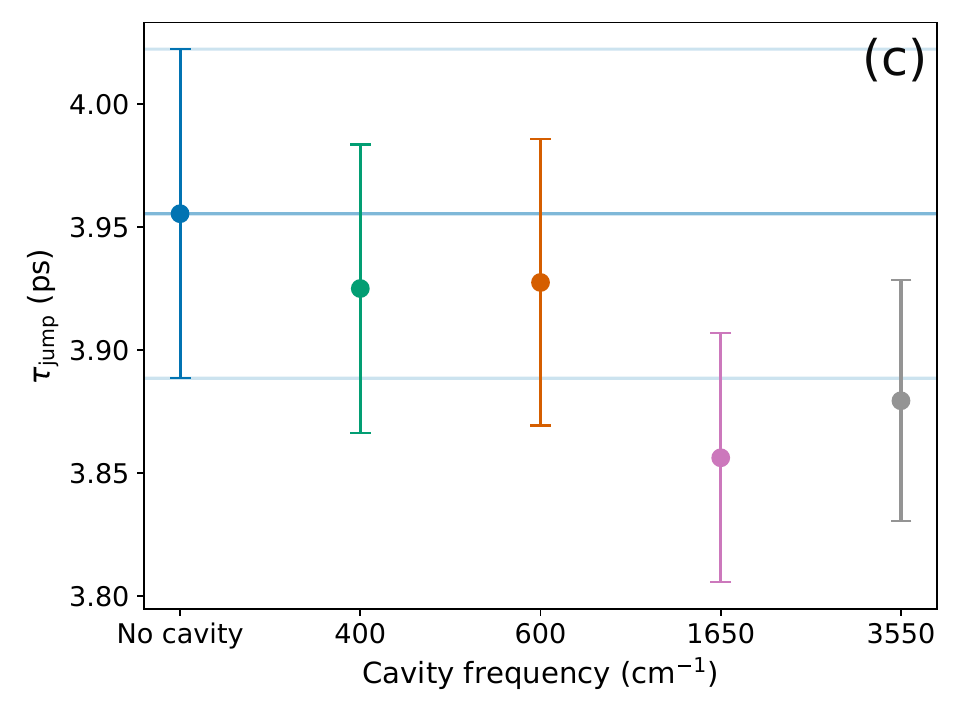}
    \caption{(a) Diffusion coefficients,   (b) OH reorientation  times (c) H-bond jump characteristic times obtained from CavMD simulations as a function of cavity frequency. The error bars correspond to 95\% of interval of confidence. The value outside cavity and corresponding error bars are reported as horizontal lines. }
    \label{fig:D_30Trajs}
\end{figure}

\begin{table*}
  \caption{\label{tbl:D_tau}For different values of cavity frequency, the HB exchange free energy barrier ($\Delta G^{\ddagger}$), the dielectric constant, the diffusion coefficient ($D$), the reorientation time ($\tau_{reor}$) and the jump time ($\tau_{jump}$) are given as obtained from CavMD simulations with classical nuclear dynamics. The reported uncertainty is  the 95\% confidence interval. }
  \begin{ruledtabular}
  \begin{tabular}{ccccccc} 
    $\omega_C$ & Coupling  &    $\Delta G^{\ddagger}$& $\varepsilon$&  $D$  & $\tau_{reor}$ & $\tau_{jump}$ \\
     (cm$^{-1}$)&(a.u.) & (kcal/mol) && (10$^{-3}$ Å$^2 \cdot $ps$^{-1}$)& (ps)& (ps) \\\hline
     No cavity & 0.000&  1.22  $\pm$ 0.03 &56 $\pm$ 2& 195 $\pm$ 6  & 3.28  $\pm$ 0.08& 3.95  $\pm$ 0.07 \\ 
               &      &                  & 52.7 $\pm$ 6.5\footnotemark[1] \\
               &      &                  & 54.5\footnotemark[2] \\ \hline
     400 
         & 4.5$\cdot$10$^{-5}$ & 1.21 $\pm$ 0.02 & 55 $\pm$ 2 &196 $\pm$ 5 &3.25 $\pm$ 0.07 & 3.92 $\pm$ 0.06\\\hline
    600  & 6.8$\cdot$10$^{-5}$ & 1.21 $\pm$ 0.03 & 57 $\pm$ 2 & 197 $\pm$ 5  & 3.22 $\pm$ 0.07 & 3.93 $\pm$ 0.06\\\hline
    1650 & 1.9$\cdot$10$^{-4}$ & 1.20 $\pm$ 0.02  & 55 $\pm$ 2 & 196 $\pm$ 5 & 3.19 $\pm$ 0.07 & 3.86 $\pm$ 0.05 \\\hline
    3550 & 4.0$\cdot$10$^{-4}$ & 1.20 $\pm$ 0.02 & 56 $\pm$ 2 &  194 $\pm$ 5  & 3.22 $\pm$ 0.05 & 3.88 $\pm$ 0.05\\
         &                   &                  & 54.5\footnotemark[2] \\
       \end{tabular}\end{ruledtabular}
       \footnotetext[1]{From Eltareb et al.~\cite{eltareb2021nuclear}}
       \footnotetext[2]{From Li et al.~\cite{li2022quantum}}
\end{table*}

We first consider the diffusion coefficients, shown in Figure~\ref{fig:D_30Trajs}(a) and Table \ref{tbl:D_tau} as a function of the cavity frequency. The value without the cavity, and the corresponding error bars, are shown as horizontal lines to help guide the eye when comparing to with the cavity. 
In panel (b) of the same Figure~\ref{fig:D_30Trajs}, the water reorientation time is represented and corresponding values are reported Table \ref{tbl:D_tau}.
Finally, in panel (c),  we report the jump time
obtained from the same simulations (again the corresponding values are listed in Table \ref{tbl:D_tau}). 

As one can see from the Figure and in details from the specific values, there is no measurable effect from the cavity for the three properties considered. It is important to notice (and we will discuss more this aspect later) that the error bars are crucial quantities to be 
carefully considered (and here we report the 95~$\%$ confidence interval). One could notice that for $\omega_c$~=~1650~cm$^{-1}$
$\tau_{reor}$ and $\tau_{jump}$ are slightly slower than when the cavity is absent.
Even if the average value is outside the error bars of the results obtained without the cavity, the difference is very small and
the interval of confidence overlap. 
As we will show in next subsection, a more detailed
statistical investigation confirms that the deviation is statistically irrelevant. 

\begin{figure}
    \centering
    \includegraphics[width=1\linewidth]{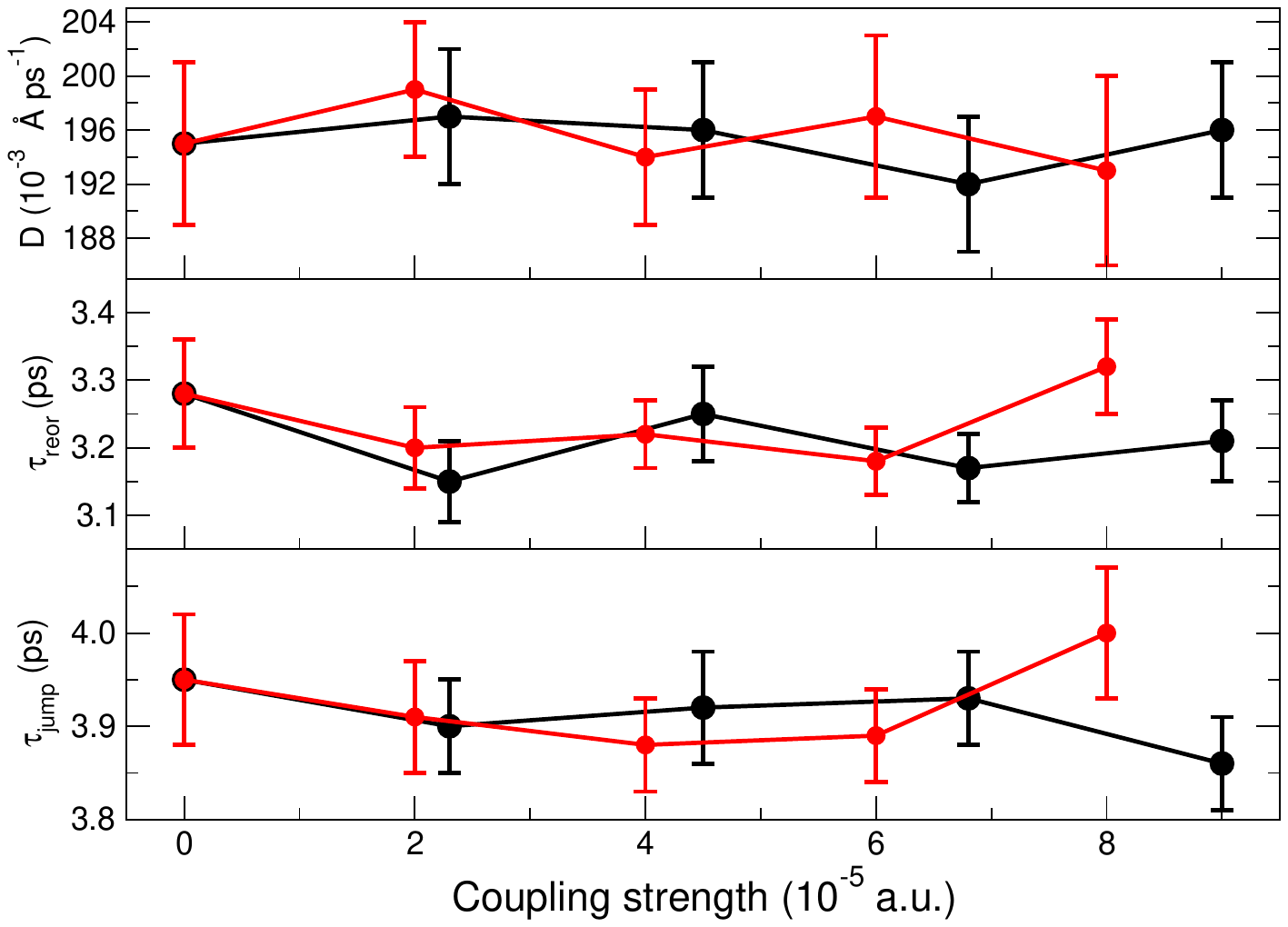} 
    \caption{Diffusion coefficient (D), water reorientation time ($\tau_{reo}$) and water jump time ($\tau_{jump}$) as 
    a function of coupling strength with the cavity for two values of cavity frequency: 400~cm~$^{-1}$ (black) and 3550~cm~$^{-1}$ (red).
    Error bars correspond to 95$\%$ of interval of confidence.} 
    \label{fig:Strength}
\end{figure}

Furthermore, even when increasing the coupling strength, we do not see any variation of such properties. In Figure~\ref{fig:Strength}, we
report results obtained when tuning the cavity at 400 and 3550~cm~$^{-1}$ at different coupling strengths: variations are inside the 
uncertainty interval and they do not show any clear trend which could be a sign of some possible effect. 

\riccardo{}
Finally, we have tested if the multi-cavity description could have an impact on such properties, with a fundamental frequency
of the cavity at 600~cm$^{-1}$.
Notably, for the 6.8~$\cdot$~10$^{-5}$ a.u coupling strength we obtained D~=~0.195~$\pm$~0.030~$\mbox{\AA}$~ps$^{-1}$, 
$\tau_{reor}$~=~3.26~$\pm$~0.23~ps and $\tau_{jump}$~=~3.97~$\pm$~0.24~ps, 
while for the coupling strength of 2.0~$\cdot$~10$^{-4}$ a.u., we got  D~=~0.209~$\pm$~0.018~$\mbox{\AA}$~ps$^{-1}$, 
$\tau_{reor}$~=~3.04~$\pm$~0.24~ps and $\tau_{jump}$~=~3.67~$\pm$~0.26~ps. These values are, within the uncertainty, equivalent to what obtained 
without the cavity or with a single cavity mode.


\subsection{Statistical Discussion}

As pointed out previously, the evaluation of statistical uncertainty is a key aspect to decipher if a property can be considered modified or not. It is also important to evaluate how many trajectories are necessary to have a reasonable convergence. 
At this end, we have employed the bootstrapping method, increasing the number of trajectories up to 60 in evaluating 
transport and dynamical properties (D, $\tau_{reo}$ and $\tau_{jump}$).
Using bootstrapping, a dataset of size $m$ with $m \in [2, 60]$ is randomly selected from the extended dataset, and the mean of this sample of size $m$ is then calculated. For each $m$, this random selection is repeated 1000 times from which we can determine an error reported here as the 95\% confidence interval.
As discussed previously, one could argue that for $\omega_c$~=~1650~cm~$^{-1}$ there can be some effect of the cavity: we have thus considered this cavity frequency in the bootstrapping analysis. 

\begin{figure}
\centering
\includegraphics[width=.9\linewidth]{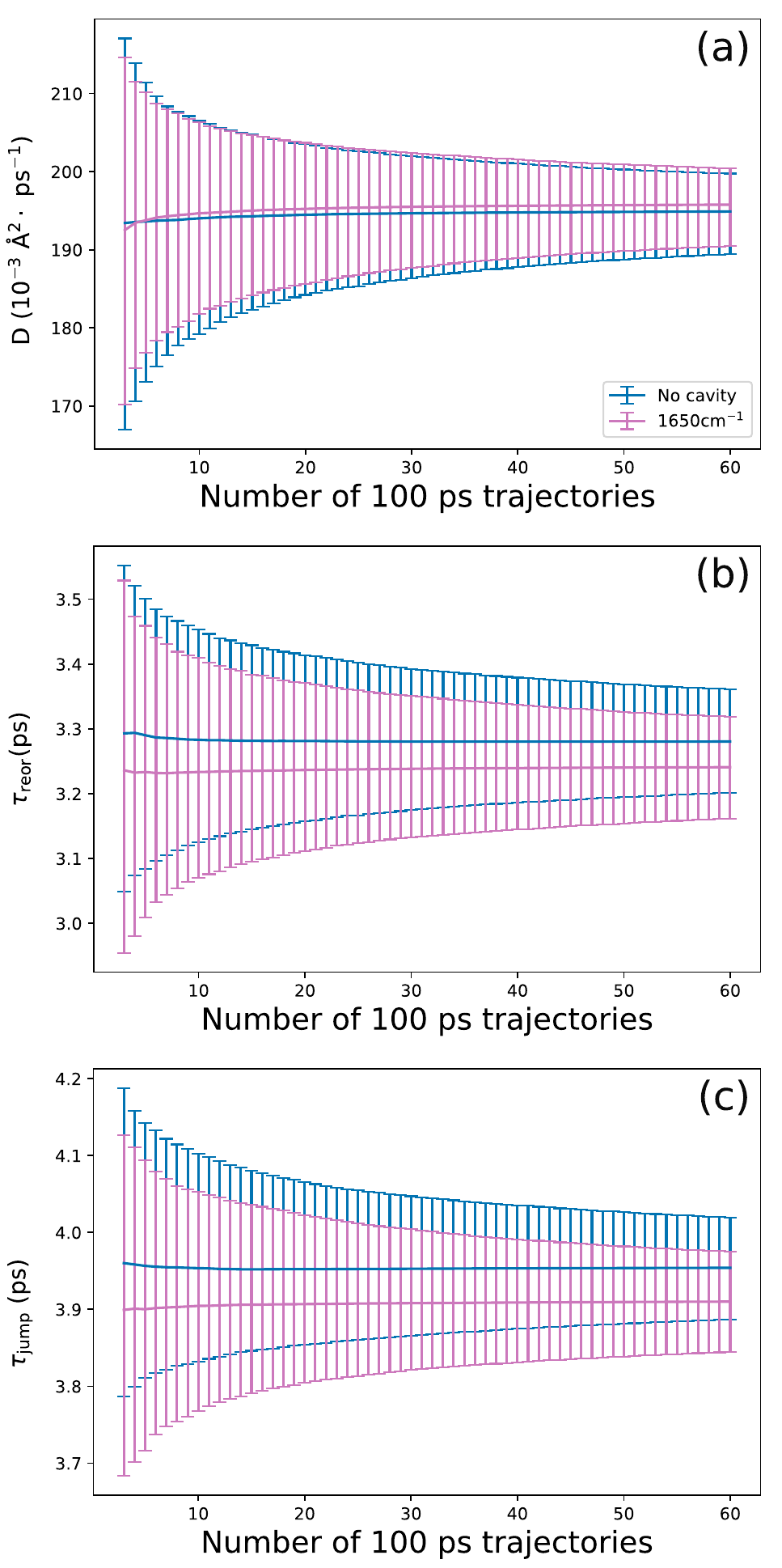}
    \caption{Mean and 95\% interval of confidence of the (a) Diffusion coefficients,   (b) OH reorientation  times (c) H-bond jump characteristic times obtained 
    for a rising number of  CavMD simulations using bootstrapping to randomly sample the data set of 60 trajectories without the cavity and for $\omega_c$=1650~cm$^{-1}$.}
    \label{fig:ConvergenceN60_1650}
\end{figure}

In Figure~\ref{fig:ConvergenceN60_1650}, we report the results as a function of trajectories, where the mean and 95\% confidence interval
are reported as a function of trajectories considered. Note that for each value a random re-sampling is performed, such that values are not
exactly the same as in Table~\ref{tbl:D_tau}, where only the (first) full set of 30~trajectories is considered. 
The results from simulations without the cavity are compared to simulations with a cavity frequency of 
1650~cm$^{-1}$. 

As shown in Figure~\ref{fig:ConvergenceN60_1650}, the cavity has no effect even increasing the number of trajectories to 60. Notably
also the size of the error bars does not change dramatically from 30 to 60. This is important, since it will allow us
to use 30 trajectories to fully converge results and to estimate a reasonable uncertainty when using less trajectories (which can 
be necessary when the computational cost increases). 

\begin{figure}
    \centering
    \includegraphics[width=.9\linewidth]{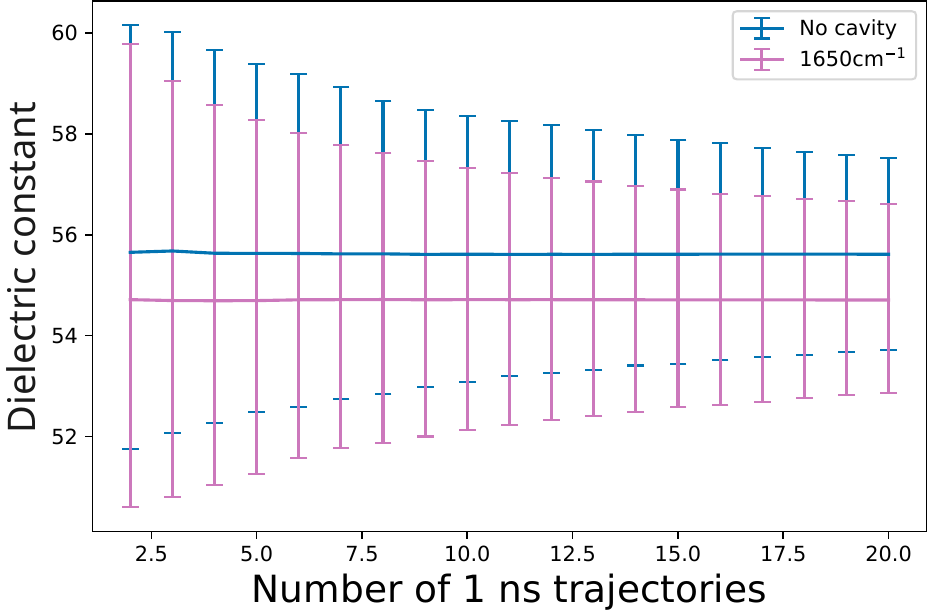}
    \caption{Mean and 95\% interval of confidence of the dieelctric constant obtained 
    for a rising number of  CavMD simulations using bootstrapping to randomly sample the data set of 20 trajectories for $\omega_c$=1650~cm$^{-1}$.}
    \label{fig:ConvergenceN60_1650_DC}
\end{figure}

The same approach is then considered for the dielectric constant, where now we consider a maximum of 20 trajectories of 1~ns each. 
The comparison between with and without the cavity (again $\omega_c$~=~1650~cm~$^{-1}$) is shown in Figure~\ref{fig:ConvergenceN60_1650_DC}. Here we show that 15 simulations is sufficient for convergence and further simulations would not change the results dramatically.

Finally, the error bars can be estimated for different number of trajectories and used for simulations where very few trajectories are 
used, which is typically the case when they are computationally expensive as is the case of RPMD with 32 beads. In the present case, we
could use the error bars estimated from the bootstrapping procedure to obtain an interval of confidence when performing only two trajectories. 




\subsection{Nuclear Quantum Effects}

As we have shown, Newtonian Born-Oppenheimer MD simulations of liquid water do not show any remarkable effect of the cavity on both 
equilibrium, transport and dynamical properties. As we have discussed (and demonstrated in the Appendix~\ref{Appendix:Inva-Equi-Prop}), the equilibrium thermodynamics properties should 
not modified via the interaction with a cavity when using the Pauli-Fierz Hamiltonian. This is not the case for 
dynamical properties, but our study shows that, at least for the system and the properties investigated, there is no effect of the cavity
neither for them. However, this is not necessarily true when considering nuclei as quantum particles. A recent study has pointed out that the inclusion of the cavity can introduce a modification of the barrier  due to a difference in zero-point
energy.~\cite{sun2024theoretical} 
Cavity Born-Oppenheimer molecular dynamics can include nuclear quantum effects (NQE) through ring-polymer MD (RPMD).~\cite{li2022quantum} These simulations
are much more expensive than classical MD, and this can be an issue given the required statistical sampling (see previous subsection). We have thus
considered RPMD with a limited number of beads (P~=~8) to investigate if NQE are relevant and tested qualitatively on few simulations with a 
larger number of beads (P~=~32).

We first compare our RPMD simulations results with P~=~8 with previous RPMD simulations in which a larger number of beads was used. Results are summarized in Table~\ref{tbl:D_tauRPMD}. 
Notably, Eltareb et al.~\cite{eltareb2021nuclear} report values of dielectric constant
and diffusion coefficient which are consistent with our results. In particular we have a very good agreement for the dielectric constant. The diffusion coefficient found in our simulations is slightly larger than those reported by them, and increasing from classical dynamics obtained in the present study. However, simulations reported by Eltareb et al. have a different number of water molecules (512) and the diffusion coefficient
was corrected for the box finite box site effect. 
While this difference can be attributed to the different number of beads and statistics, a similar behaviour (i.e. the increasing of diffusion coefficient from classical to RPMD simulations) was 
obtained by Habershon et al.~\cite{habershon2009competing}. 

We have then turned the cavity on at the 600~cm~$^{-1}$ frequency, which corresponds to librational motion also when adding NQEs. 
Also in this case we do not observe any effect of the cavity on both dielectric constant and diffusion coefficient as
reported in Table~\ref{tbl:D_tauRPMD}. 
We  notice that Li et al. have also found no effect of 
the cavity on dielectric constant when tuning the cavity 
at the O--H stretching frequency.~\cite{li2022quantum}

%

Moving to the water reorientation and jump times, Wilkins et al. determined a small acceleration when moving from classical to RPMD simulations.~\cite{wilkins2017nuclear} Our simulations show the same behaviour. When then adding the cavity with RPMD simulations, these two dynamical quantities are also unchanged as obtained in Newtonian simulations. Note that the 
cavity is tuned to 600~cm$^{-1}$ as this lower frequency of the librational modes contributes towards the  mechanism responsible for H-bond jumps through the angular motion of the O--H bond.~\cite{laage2006molecular,laage2008molecular}


\begin{table}
  \caption{\label{tbl:D_tauRPMD} Results obtained from RPMD simulations with P~=~8 without the cavity and with the cavity set at 600~cm~$^{-1}$ with a coupling strength of 6.8~$\cdot$~10~$^{-5}$ au. Uncertainties are given as 95$\%$ of confidence interval. \riccardo{We also report results
  obtained by Li et al.~\cite{li2022quantum} when setting the cavity resonant with O--H stretching and using P~=~32.}}
  \begin{ruledtabular}
  \begin{tabular}{l | ccc}
    & No Cavity & $\omega_c$~=~600~cm~$^{-1}$ & $\omega_c$~=~3400~cm~$^{-1}$\\
    \hline
$\epsilon$ &     55 $\pm$ 2 & 58 $\pm$ 2 \\
           &     53 $\pm$ 7\footnotemark[1] & \\
           &     59\footnotemark[3] & & 56.6\footnotemark[3]\\
$D$ (10$^{-3}$ Å$^2 \cdot $ps$^{-1}$) & 246 $\pm$ 6 & 247 $\pm$ 7 \\
                                     & 220\footnotemark[1] & \\
                                    & 221\footnotemark[2] $\pm$ 1 & \\
$\tau_{reor}$ (ps) & 2.54 $\pm$ 0.07 & 2.51 $\pm$ 0.07 \\
$\tau_{jump}$ (ps) & 3.13 $\pm$ 0.04 & 3.13 $\pm$ 0.05 
            \end{tabular}\end{ruledtabular}
            \footnotetext[1]{From Eltareb et al.~\cite{eltareb2021nuclear}} 
            \footnotetext[2]{From Habershon et al.~\cite{habershon2009competing}}
            \footnotetext[3]{\riccardo{From Li et al.~\cite{li2022quantum}}}
\end{table}

Results with P~=~8 are reported by averaging 50 trajectories of 300~ps for the dielectric constant and from 30 trajectories  of 100~ps for the other properties. According to our statistical discussion, theses simulations times are enough to be able to give a conclusion on the results. Simulations with P~=~32 were 
done only on two trajectories of 100~ps in order to calculate diffusion coefficient, reorientation and jump times. They are estimated at 0.216~$\pm$~0.015~Å$^2 \cdot $ps$^{-1}$ , 2.7~$\pm$~0.3~ps and 3.5~$\pm$~0.3~ps (the uncertainties are estimated from the bootstrap analysis done in the classical trajectories). These results show that increasing the number of beads do not change the fact that the cavity, within this simulation method and for these properties, has no effect them.

\section{Conclusions}

In the present paper, we have studied equilibrium and dynamical properties of liquid water under vibrational strong coupling by
using the Born-Oppenheimer cavity MD method proposed by Li et al.~\cite{li2020cavity,li2022quantum} in which the q-TIP-4P/F model is
used. At this end, we have set the cavity frequency to the different modes present in the water IR spectrum, corresponding to the O--H stretching,
H--O--H bending and librational modes. 

Similarly to what observed by Li et al.~\cite{li2020cavity,li2022quantum} we do not observe any effect of the cavity on any property investigated,
also monitoring other cavity frequencies in resonance with bending or librational motion (previous works studied only the cavity in resonance with O--H stretching). We extended the analysis to other properties, with respect to what present in the literature, and namely the diffusion coefficient and dynamical properties of water motion (reorentational  and jump times). Also these properties are not modified by the presence of the cavity
at any frequency in resonance with different liquid water IR bands. 
The inclusion of nuclear quantum effects via RPMD approach does not change the picture.

We should notice, that the experimental possibility of a modification of the dielectric constant due to VSC had a controversal history in the 
recent literature: while a 2023 experimental paper reported an effect on it,~\cite{piejko2023solvent} it was retracted very recently~\cite{piejko2024Retract} and an explanation was given by
Michon and Simpkins.~\cite{Michon2024} Our simulations are in agreement with those results and our statistical analysis shows that one has to be very 
careful in statistical convergence and uncertainty evaluation also in simulations. This is true not only for dielectric constant (which needs
long trajectories to converge as it is well-known) but also for other properties, for which one can see illusory effects without a 
correct statistical analysis.
%

Concerning the diffusion coefficient, experiments have shown that the diffusion of ions can be modified by VSC and in case of water this is
likely due to a difference in proton conductivity.~\cite{fukushima2022inherent,fukushima2023} The non reactive q-TIP4P/F model clearly cannot provide any support on it and our study 
shows that with this popular water model, the diffusion coefficient is not altered by VSC. 
Unfortunately there are no experimental studies on the H-bond reorentational times under VSC.
H-bond jump times are not accessible experimentally, but were studied here  since they hold all the properties of a chemical reaction
even if no covalent bond is broken/made, but only H-bonds. 
Also on those properties related to H-bond dynamics, using the present model, we cannot highlight any effect of VSC at any cavity frequency.

As pointed out, we have extensively studied liquid water under VSC employing the q-TIP4P/F water model. This model has two limitations which could 
be at the origin of the absence of any relevant effect of VSC on dynamical properties:  (i) it is a fixed charge model (the cavity field does not change the charge
distribution of water molecules); (ii) the water model does not allow bond breaking/making.
Nevertheless, the model is able to recover the spectroscopic features, not only in terms of formation
of polaritonic bands in the IR signal but also the small blue and red shifts of other bands not in resonance with the cavity mode. 
\riccardo{As pointed out by Lieberherr et al.~\cite{lieberherr2023vibrational}, the coupling strength used to reproduce the typical Rabi splitting using the q-TIP4P/F water model is unphysically high. This provides additional support to the conclusions of our work : since an extremely large coupling strength does not lead to any significant impact on any of the investigated water properties, one can safely suggest that lower coupling values should not cause any meaningful change in the liquid water properties. In addition, we note that while the 
qTIP4P-F model already provides an excellent description of the properties of water, future extensions of this work will examine VSC effects with more sophisticated water models.}

The fixed cavity model could be improved in the future considering molecular polarizability, as a first step.
In fact,  recent theoretical studies by Rubio and co-workers have shown that VSC could modify the polarizability of molecules present in 
the cavity.~\cite{sidler2024unraveling} 
Unfortunately, their method is based on very expensive calculations which cannot be presently applied to liquid systems. 
Water models including polarizability can be employed as a first approximated description of fluctuating molecular electronic density (via
 induced dipoles, for example)
and the Pauli-Fierz Hamiltonian can be adapted.
Additionally, an interesting question is whether the inclusion of polarization, in classical MD, through models such as the ﬂuctuating charges
model, the Drude model or the induced dipoles model is sufficient or is the consideration of the electronic structure necessary.
In any case, it will be important to both consider a number of molecules under VSC
which is sufficiently large (at least 100)~\cite{JaimeAccepted} and have a statistical convergence of the properties. 
Our research are now going in that direction.

\begin{acknowledgments}
The authors thank Dr. Tao Li for providing support in  utilising the CavMD software.
This work was supported by the French National Research Agency ANR (ANR-20-CE29-0008-02, MoMaVSC).
We also thank IDRIS for generous allocation of computing time (project 103847).
\end{acknowledgments}

\section*{Author Declarations}
\subsection*{Conflict of Interest}
The authors have no conflicts to disclose.







\section*{Data Availability Statement}

The data that support the findings of this study are available from the corresponding author upon reasonable request.

\appendix

\section{3-Mode Model Pauli--Fierz Hamiltonian and Spectra Calculations from Path Integral Monte Carlo}\label{Appendix:PIMC}

We have considered the Pauli-Fierz Hamiltonian, Equation~\ref{eq:PF_ham}, where now
the matter potential is expressed via three harmonic oscillators centered at $\omega_{lib}$~=~650~cm~$^{-1}$, $\omega_{bend}$~=~1650~cm~$^{-1}$ and
$\omega_{str}$~=~3500~cm~$^{-1}$: 

\begin{equation}
V_{M}= \frac{1}{2} m \sum_{i=1}^{N_{lib}} \sum_{j=1}^{N_{bend}} \sum_{k=1}^{N_{str}} \left( \omega_{lib,i}^2 q_{lib,i}^2+ \omega_{bend,j}^2 q_{bend,j}^2 + \omega_{str,k}^2 q_{str,k}^2 \right) 
\end{equation}

where for each mode we have sampled a Gaussian distribution \riccardo{such that the integral of the cavity-free signal obtained from simulations
is obtained at best. Here we set $N_{lib} + N_{lib} + N_{str} = 100$. This finite distribution is at the origin of the small discrepancies on the
cavity-free spectrum which in any case do not modify the main results.} The coupling constant in the Pauli-Fierz Hamiltonian is empirically set to reproduce the Rabi splitting observed in the present CavMD simulations.

With this Hamiltonian, we set the cavity at 1650~cm~$^{-1}$ and
run Path Integral Monte Carlo (PIMC) simulations from which vibrational frequencies are obtained. Details are described in our
recent work~\cite{JaimeAccepted}. Here we give a short overview.

Specifically, we run 10$^8$ PIMC steps at 300~K using 64 beads and then we averaged over 50 replicas. From these simulations we calculated the displacement-displacement correlation function in the Path Integral formalism and solved the generalized eigenvalue
problem following Morresi et al.~\cite{Morresi2021} 
In this way we obtained the transition frequencies and 
the IR intensity is obtained from the matter components of the eigenvectors as detailed in Ref.~\cite{JaimeAccepted}


\section{Invariance of Structural and Thermodynamic Properties for Classical Particles under VSC}\label{Appendix:Inva-Equi-Prop}

In this section, we will show how, as long as a classical canonical statistical mechanics are considered, the structural or thermodynamic properties of matter remain invariant under VSC, described by the Pauli\--Fierz Hamiltonian in Eq.\eqref{eq:PF_ham}.
\riccardo{A concise derivation is given by Li et al.~\cite{li2020cavity}}\dl{ and here we provide an extended version of it.} 

The structural or thermodynamic value of an arbitrary property A, well defined at every point in phase space, can be computed by:

\begin{eqnarray}    
   \langle A\rangle &=& \int_{-\infty}^\infty dp_1\hdots\int_{-\infty}^\infty dq_1\hdots P\left(q_1,q_2,\hdots q_{3N}; p_1,p_2,\hdots, p_{3N}\right) \nonumber \\
   & & A\left(q_1,q_2\hdots q_{3N}; p_1,p_2,\hdots, p_{3N}\right)
\end{eqnarray}

Where $P(\left\{p_i,q_i\right\})$ is the probability of visiting the point in phase space given by the phase space vector $\left(q_1,q_2\hdots q_{3N}; p_1,p_2,\hdots, p_{3N}\right)$. We will, for the sake of compactness, from now on abbreviate a phase space vector as $\left(\mathbf{q},\mathbf{p}\right)$. We will show that the phase space distribution function for matter is not affected by coupling to the cavity. To do so, consider that in the canonical ensemble, the solution to the Liouville equation that gives the phase space distribution function is:

\begin{equation}
    P\left(\mathbf{q},\mathbf{p}\right) = \frac{e^{-\beta H}}{Z_T}
\end{equation}

Where $\beta = 1/k_BT$, and $Z_T$ is the partition function over all degrees of freedom, defined as $Z_T= \int_{-\infty}^\infty d\mathbf{p}\int^\infty_{-\infty}d\mathbf{q} e^{-\beta H(\mathbf{q},\mathbf{p})}$. Let us now consider separately the degrees of freedom of the cavity, $\mathbf{q}_c$ and $\mathbf{p}_c$, which are the field coordinates and it conjugated momenta, respectively, for all the field modes and polarizations involved in the model. We also denote collectively $\mathbf{q}_M$ and $\mathbf{p}_M$ all the degrees of freedom relating to the matter part. If one is interested in an structural or thermodynamic property which is defined only on the matter part of the system, one must compute the marginal probability density function associated with the matter subsystem by integrating over all cavity degrees of freedom.
\begin{equation}
    P(\mathbf{q}_M,\mathbf{p}_M) =\frac{1}{Z_T}\int^\infty_{-\infty} d\mathbf{q}_c\int^\infty_{-\infty} d\mathbf{p}_c e^{-\beta H}
    \label{eq:marginal_probability}
\end{equation}
In order to perform the integration, we note that the Hamiltonian in Equation \eqref{eq:PF_ham} can be written, in terms of our collective notation for the modes, as:

\begin{equation}
    H = \sum_{k,\lambda} \left[\frac{1}{2}p_{k,\lambda}^2 + \frac{1}{2}\omega^2_{k,\lambda}\left(q_c+ f(\mathbf{q}_M,\mathbf{p}_M)\right)^2\right] + H_M\left(\mathbf{q}_M,\mathbf{p}_M\right)
    \label{eq:summarized_pf}
\end{equation}

Where we stress that the last term in the parenthesis is only a function of matter degrees of freedom, and $H_M$ is the Hamiltonian that describes the isolated matter subsystem. Now, going back to the integral in Eq\eqref{eq:marginal_probability}, as we restrict ourselves to classical statistical mechanics, we can factorize the sum over modes and polarizations in a product of exponentials:

\begin{eqnarray}
P(\mathbf{q}_M,\mathbf{p}_M) &=& \frac{e^{-\beta H_M(\mathbf{q}_M,\mathbf{p}_M)}}{Z_T} \int_{-\infty}^\infty dp_{1,1}e^{-\frac{\beta}{2}p^2_{1,1}} \nonumber \\
& & \int_{-\infty}^\infty dp_{1,2}e^{-\frac{\beta}{2}p^2_{1,2}}
\int_{-\infty}^\infty dp_{2,1}e^{-\frac{\beta}{2}p^2_{2,1}}\hdots\nonumber\\
& & \int^\infty_{-\infty} dq_{1,1}e^{-\frac{\beta}{2}\omega_{1,1}\left(q_{1,1}+f(\mathbf{q}_M,\mathbf{p}_M)\right)^2} \nonumber \\
& & \int_{-\infty}^\infty dq_{1,2}e^{-\frac{\beta}{2}\omega_{1,2}\left(q_{1,2}+f(\mathbf{q}_M,\mathbf{p}_M)\right)^2} \nonumber \\
& & \int^\infty_{-\infty} dq_{2,1}e^{-\frac{\beta}{2}\omega_{2,1}\left(q_{2,1}+f(\mathbf{q}_M,\mathbf{p}_M)\right)^2}\hdots
\label{eq:factorized_exponential}
\end{eqnarray}

We can here recognize that each integral involving a field momentum is a gaussian integral, that is solved by $\int^\infty_{-\infty}dxe^{-ax^2}=\sqrt{\frac{\pi}{a}}$. If we let $M$ be the number of modes considered, and 2 polarizations are allowed for each mode:

\begin{align}
    P(\mathbf{q}_M,\mathbf{p}_M) =& \left(\frac{2\pi}{\beta}\right)^M\frac{e^{-\beta H_M(\mathbf{q}_M,\mathbf{p}_M)}}{Z_T}\int dq_{1,1}e^{-\frac{\beta}{2}\omega_{1,1}\left(q_{1,1}+f(\mathbf{q}_M,\mathbf{p}_M)\right)^2}\nonumber\\
    &\int_{-\infty}^\infty dq_{1,2}e^{-\frac{\beta}{2}\omega_{1,2}\left(q_{1,2}+f(\mathbf{q}_M,\mathbf{p}_M)\right)^2} \nonumber \\
    & \int^\infty_{-\infty} dq_{2,1}e^{-\frac{\beta}{2}\omega_{2,1}\left(q_{2,1}+f(\mathbf{q}_M,\mathbf{p}_M)\right)^2}\hdots\nonumber\\
    \label{eq:momenta_integrated}
\end{align}
A second set of gaussian integrals can also be recognized, since the terms involving the exponential of the field coordinate involve a term dependent exclusively on the matter coordinates, which is a constant to the integral. A change of variable of the form $x = q+f(\mathbf{q}_M,\mathbf{p}_M)$ allows to integrate each cavity coordinate term:

\begin{eqnarray}    
P(\mathbf{q}_M,\mathbf{p}_M) &=& \left(\frac{2\pi}{\beta}\right)^{2M}\left(\frac{1}{\prod_{k,\lambda}\omega^2_{k,\lambda}}\right)^{1/2}\frac{e^{-\beta H_M(\mathbf{q}_M,\mathbf{p}_M)}}{Z_T} \nonumber \\
&=& \frac{e^{-\beta H_M(\mathbf{q}_M,\mathbf{p}_M)}}{Z_M} 
    \label{eq:all_integrated}
\end{eqnarray}

Where $Z_M$ is the partition function of a system consisting exclusively on matter, $Z_M = \int d\mathbf{q}_M\int d\mathbf{p}_M e^{-\beta H_M(\mathbf{q}_M,\mathbf{p}_M)}$. All contributions to the cavity have been integrated, and all have yielded constant factors, that do not favor any particular matter configuration. The phase space distribution inside the matter manifold is controlled solely by $e^{-\beta H_M}$, as it would be in the case of a ordinary system without interaction to field variables. Therefore, this shows that within the Pauli\--Fierz Hamiltonian description, VSC does not affect the structural or thermodynamic values of matter\--only properties following classical statistical mechanics. 
\bibliography{aipsamp}

\end{document}